# PEG Branched Polymer for Functionalization of Nanomaterials with Ultralong Blood Circulation

*Giuseppe Prencipe, Scott M. Tabakman, Kevin Welsher, Zhuang Liu, Andrew P. Goodwin, Li Zhang, Joy Henry and Hongjie Dai\**

\* Department of Chemistry Stanford University, Keck Science Building, Rm 125 380 Roth Way

Stanford, CA 94305

Phone: 650-723-4518 (office); 650-725-9156 (lab)

Fax: 650-725-0259

Email: hdai or hdai1@stanford.edu

Office: Keck Science Building, Room 125

Website: http://www.stanford.edu/dept/chemistry/faculty/dai/group/

*Abstract:* Nanomaterials have been actively pursued for biological and medical applications in recent years. Here, we report the synthesis of several new poly(ethylene glycol) grafted branched-polymers for functionalization of various nanomaterials including carbon nanotubes, gold nanoparticles (NP) and gold nanorods (NRs), affording high aqueous solubility and stability for these materials. We synthesize different surfactant polymers based upon poly-(γ-glutamic acid) (γPGA) and poly(maleic anhydride-*alt*-1-octadecene) (PMHC$_{18}$). We use the abundant free carboxylic acid groups of γPGA for attaching lipophilic species such as pyrene or phospholipid, which bind to nanomaterials via robust physisorption. Additionally, the remaining carboxylic acids on γPGA or the amine-reactive anhydrides of PMHC$_{18}$ are then PEGylated, providing extended hydrophilic groups, affording polymeric amphiphiles. We show that single-walled carbon nanotubes (SWNTs), Au NPs and NRs functionalized by the polymers exhibit high stability in aqueous solutions at different pH's, at elevated temperatures and in serum. Morever, the polymer-coated SWNTs exhibit remarkably long blood circulation ($t_{1/2}$ 22.1 h) upon intravenous injection into mice, far exceeding the previous record of 5.4 h. The ultra-long blood circulation time suggests greatly delayed clearance of nanomaterials by the reticuloendothelial system (RES) of mice, a highly desired property for in vivo applications of nanomaterials, including imaging and drug delivery.

**Introduction**

Nanostructures including SWNTs, Au NPs and Au NRs are currently being explored for biomedical applications, including in vivo delivery (e.g. drugs,[1, 2] proteins, peptides[3-5] and nucleic acids[2, 4-9]), biological detection of proteins[10-12] and in vivo imaging[13]. Most inorganic nanomaterials are not soluble in physiological buffers, and require functionalization by thiols or surfactants to obtain biocompatibility. Many surfactant coatings become unstable in serum, or upon removing the excess of the coating molecule. PEGylation is a common strategy to impart functionality, water solubility and biocompatibility[14-17]. Even so, in vivo, most of the existing functionalization methods of nanomaterials suffer from rapid uptake by the reticuloendotheial system (RES) and short blood circulation times. Sufficient blood circulation time is critical to both imaging and in vivo delivery.

In recent years, advances have been made toward the use of polymeric amphiphiles to coat nanomaterials non-covalently. For SWNTs, non-covalent functionalization is necessary to preserve the intrinsic physical properties of SWNTs, including near-infrared fluorescence and Raman scattering, useful for biomedical imaging[18]. To this end SWNTs have been incorporated into micelles formed by various surfactants in aqueous solutions, such as lipids[19, 20], sugars[21], proteins,[22, 23] DNA,[24, 25] and polymers like polyethylene glycol (PEG)[26, 27]. Also, it has been found that densely coating nanomaterial surfaces with PEG[28, 29] increases in vivo circulation times, likely by resisting clearance via the RES. Still, much room exists to optimize non-covalent nanomaterial surface coating, improving the circulation behavior of SWNTs for in vivo biomedical applications including tumor targeting, imaging and therapy.

Similarly to SWNTs, inorganic colloidal nanoparticles, such as gold nanoparticles (NP) or gold nanorods, have been extensively investigated for two decades, and have found application largely in detection schemes for DNA,[7-9] proteins,[10, 11] and other biomolecular analytes. NPs and NRs have also been used in single-particle coding and tracking experiments in vitro,[30-35] and in vivo.[36] These applications require robust, aqueous nanoparticle suspensions, which have relied upon surface passivation by small molecules,[37-42] lipids,[43-45] surface silanization[46-49] and amphiphilic polymer coatings. There is significant ongoing research regarding the synthesis of stabilized Au NPs in the presence or in the absence of thiol ligands.[50-52] Pyrene-containing and phospholipid-coating moieties have been used extensively for binding to carbon nanotubes[13, 28, 53, 54] and gold nanoparticles[55], due to strong interactions including van der Waals forces, π-π stacking, charge transfer, and/or hydrophobic interactions [28, 56, 57].

Here, we report the synthesis of several new poly(ethylene glycol) grafted branched-polymers based upon poly-(γ-glutamic acid) (γPGA) and poly(maleic anhydride-*alt*-1-octadecene) (PMHC$_{18}$). Poly-γ-glutamic acid (γPGA) is a naturally occurring bio-material, produced by microbial fermentation[58]. γPGA is water soluble, biodegradable, nontoxic and edible. We propose that a portion of the free carboxylic acids of γPGA could first be coupled to lipophilic groups for binding to nanomaterials via hydrophobic and van der Waals interactions, while the remaining carboxylic acids can be conjugated to PEG, providing enhanced aqueous solubility and further biocompatibility[59]. In the case of **1**, the copolymer itself contains an alternating hydrophobic unit, the C$_{18}$ chain. Specifically, maleic anhydride groups may be reacted with primary amine-terminated PEG, and remaining carboxylic acids can be conjugated to PEG via amidation chemistry.

We found that not only was this material able to form self-assemblies easily in water, but also formed stable coatings on carbon nanotubes, gold nanoparticles, and gold nanorods. These polymer-nanomaterial coatings showed stability to a range of pH's, salt conditions, and introduction of serum. Finally, polymer-coated SWNTs exhibit remarkably long blood circulation ($t_{1/2}$ 22.1 h) upon intravenous injection into mice, when compared with the previous record of 5.4 h.

**Results and discussion**

**Synthesis of polymers.** Pyrene and phospholipid were chosen as lipophilic groups for attachment to biocompatible polymers. In the first synthetic step (see Supporting Information for methods), we used the free carboxylic acid of γPGA ($M_n$ ~ 430 KDa) to couple 1-methylaminopyrene via EDC amidation. In a second step, we used the remaining carboxylic acid groups of γPGA to attach amine-terminated poly(ethylene glycol) methyl ethers (mPEG-$NH_2$, MW 5000) to afford the compound (**2**). The average number of pyrene moieties grafted to γPGA chains was determined by $^1$H NMR estimated to be ~30% and ~70% of γPGA backbone was loaded with PEG chains. The lipophilic pyrene moiety was then replaced by the biocompatible phospholipid 1,2-distearoyl-*sn*-glycero-3-phosphoethanolamine (DSPE) by first PEGylating γPGA and then grafting the lipophilic phopholipid moiety in the final step (see Supporting Information for methods) to afford the compound (**3**). The average numbers of DSPE and PEG moieties grafted to γPGA chains were determined by $^1$H NMR, to be ~10% and ~60% respectively.

We also synthesized a PEGylated amphiphilic polymer based on a poly(maleic anhydride-*alt*-1-octadecene) ($PMHC_{18}$) backbone. In this case the copolymer itself contains an alternating hydrophobic unit, the $C_{18}$ chain. Specifically, in the first synthetic step (see Supporting Information for methods), maleic anhydride groups were reacted with primary amine-terminated poly(ethylene glycol) methyl ethers (mPEG-$NH_2$). The remaining carboxylic acids of $PMHC_{18}$ were coupled to additional mPEG-$NH_2$ via EDC amidation to obtain a fully PEGylated, highly water soluble amphiphile polymer (**1**). The average number of PEG moieties grafted to PMHC$_{18}$ chains was determined by $^1$H NMR, to be ~ 200% (2 molecules of mPEG per unit of PMHC$_{18}$).

All three polymers are completely soluble in both chloroform and in aqueous media. Thus, in aqueous conditions hydrophobicity-driven self-assembly would be expected. In particular, the aqueous polymers are able to form micelles. We measured the CMC (critical micelle concentration) for each polymer (see Supporting Information Fig. S1). Surface tension measurement yielded CMC's of 0.2 mg/mL for **1**, 0.3 mg/mL for **2** and 0.4 mg/mL for **3**. Also, we used dynamic light scattering to measure the size distribution of the micelles self-assembled in water (see the Supporting Information Fig. S2-S3). The surfactant molecules' mean hydrated diameters were ~17 nm, ~33 nm and ~56 nm for **1**, **2** and **3** respectively. Whereas micelles of the polymers have hydrated diameters of ~132 nm, ~232 nm and ~196 nm for **1**, **2** and **3** respectively.

**Functionalization of Carbon Nanotubes by 1-3.** Fig. 2 (see also the Supporting Information Fig. S5-S6) illustrates that following sonication in polymer solutions, we obtain excellent suspensions of SWNTs stabilized in water by polymers (**1-3**) , even after removal of excess polymer by repeated vacuum filtration (200 nm pore size). SWNT suspensions showed excellent stability, without aggregation or deviation of native UV/Vis/NIR absorbance, at pH's ranging from 1 to 12, at high temperature overnight, and in 50% fetal calf serum for 48 h. AFM images (Fig. 2b ) show mostly dispersed SWNTs with lengths of ~200 nm. The UV-Vis (Fig. 2c) absorbance spectrum of SWNTs shows van Hove singluarities typical of well-dispersed SWNTs, with characteristic $E_{11}$ and $E_{22}$ transistions. The pyrene, the $C_{18}$, and DSPE moieties have a strong tendency to adsorb on SWNTs by hydrophobic interactions in aqueous media, and, when applicable, by π-stacking. In this way, we obtain robustly coated SWNTs in aqueous media. Proof of the intimate interaction between pyrene and the NT sidewall was observed by quenching of pyrene fluorescence (see Supporting Information S7) relative to free (**2**) with the same OD. Fig. 2d shows the photoluminescence versus excitation (PLE) spectrum of SWNTs, the inherent NIR photoluminescent properties of SWNTs are retained when they are coated with (**1**).

**Functionalization of Gold Nanoparticles by 2-3.** In addition to directly suspending NTs in the case of γPGA polymers (**2** and **3**), it is possible to exchange other nanostructures from their native capping ligand into our amphiphilic γPGA polymer. For example, very good suspensions of gold NPs in water were obtained through sonication for 10 min in the presence of excess **2** or **3** to displace citrate. Excess citrate was removed by dialysis, and the excess of surfactant was removed by repeated centrifugation. This procedure gave mostly dispersed nanoparticles, as shown by TEM (Fig. 3a). As with NTs, this suspension was observed to be stable under various conditions, such as neutral-basic pH's, at 70°C overnight, and in 50% serum, showing no significant changes in suspension dispersity or absorbance after a 48 h incubation. In contrast, thiol-mPEG(5KDa), a strong and covalent passivator of gold nanoparticles, showed less stability. In particular, as shown in (Fig. 3b), the NPs-thiol-mPEG(5KDa) are stable only in the presence of excess thiol-mPEG (see the Supporting Information S9). Indeed if the excess of the thiol-mPEG is removed by centrifugation the solution of NPs becomes unstable, forming aggregates. The UV/visible spectrum (Fig. 3a) shows the absorbance of gold nanoparticles at 530 nm. In this case (see Supporting Information S8), the intensity of pyrene adsorbance is dampened due to perturbations of gold nanoparticles, suggesting direct interaction of pyrene with the gold surface.[55] As in the case of NTs, the pyrene and DSPE moieties were adsorbed on the gold suface through hydrophobic interactions.

**Functionalization of Gold Nanorods by 2.** We also used our polymeric amphiphile to suspend gold nanorods. For example, very good suspensions of gold NRs in water were obtained through sonication for 15 min in presence of excess **2** to displace CTAB (hexadecyl-trimethyl-ammonium bromide). Excess CTAB was removed by dialysis, and the excess of surfactant was removed by repeated centrifugation. This procedure gave mostly dispersed nanorods, as shown by TEM and UV/vis absorbance spectroscopy (Fig. 3c). Suspensions of gold nanorods with **2** were stable at neutral-basic pH's, at 70°C overnight, and in 50% serum for 48 h. This result is important because nanorods with covalent thiol-based passivation are unstable under similar conditions. The UV/vis absorbance spectrum in (Fig. 3c) shows the transverse and longitudinal adsorbance of gold nanorods suspended by (**2**) or by CTAB at 520 nm and 860 nm respectively. While the solutions were originally normalized to the same OD, removal of excess surfactant causes significant aggregation and consequent loss of plasmon absorbance for CTAB-suspended NRs, however NRs suspended by (**2**) remain stable, even in the presence of serum. Again, the intensity of pyrene absorbance is dampened (data not shown) by close proximity to the gold nanorods. The TEM picture shows the size distibution and dispersity of nanorods suspended with **2**.

**Circulation time of Carbon Nanotubes functionalize by 1-3.** Long blood circulation half-life of a drug carrier is desired to improve the bioavailability of the drug. For most nanomaterials this has been an elusive goal. For example, after systemic administration, carbon nanotubes are gradually cleared from the circulating blood by macrophage uptake as part of the RES, which leads to accumulation in the liver and spleen. Therefore, prolonged circulation can only be achieved if rapid RES uptake is avoided. In our previous studies we have shown that surface-coating of SWNTs can delay RES uptake and thus increase blood circulation[28, 56, 57].

Hypothesizing that the pronounced PEG loading on the polymer-SWNTs would prolong the blood circulation, we injected these conjugates into mice. Importantly, we observed (Fig. 4a) a circulation half-life of 2.4 h for SWNTs coated with (**3**), a very long circulation half-life of 22.1 h for SWNTs coated with (**2**) and 18.9 h for SWNTs coated with (**1**), which were far longer than values reported by others, as well as our previous data obtained for DSPE-coupled linear PEG coated SWNTs (0.33 h for DSPE-l-2kPEG, 2.4 h for DSPE-l-5kPEG, Fig. 4b) or DSPE-branched PEG functionalized SWNTs (5.4 h for DSPE-br-7kPEG, Fig. 4 a&b). While the packing density of PEG coatings immobilized on nanotubes via conventional surfactants with single anchoring points is limited by steric hindrance, the multiple lipophilic anchoring domains and multiple PEG chains of (**2**) and (**1**) allow continuous binding of the polymer onto the nanotube surface, yielding a highly dense PEG coating. Additionally, the molecular weight and functionalization density of each PEG branch was important for obtaining long circulation. Similar surfactants produced with mPEG 0.75 kDa in lieu of 5 kDa led to significantly reduced circulation times. Moreover, functionalization of the γPGA backbone with less than 50% of mPEG 5KDa gave unsatisfactory circulation results. Thus, not only does this unique polymer coating hold promise for potential in vivo therapeutic applications, we have also been able to discern what aspects of the polymer lead to desirable properties in our SWNT conjugates.

**Conclusion**

In conclusion, we report the synthesis of new versatile PEGylated branched polymers that provide stable suspensions of carbon nanotubes, gold nanoparticles and gold nanorods. We used poly-γ-glutamic acid (γPGA) functionalized with either pyrene moieties or phospholipids to provide robust polymer-particle interactions and PEG, enhancing aqueous solubility and biocompatibility. We also used PEGylated poly(maleic anhydride-*alt*-1-octadecene) (PMHC$_{18}$) to provide very stable suspensions of carbon nanotubes. We showed that the SWNTs, Au NPs and Au NRs functionalized by these polymers show excellent stability in aqueous solutions, at different pH's, and elevated temperatures, as well as in serum. Moreover, the polymer-coated SWNTs exhibit unprecedentedly long blood circulation (half-life 22.1 h for (**2**) and 18.9 h for (**1**)) upon intravenous injection into mice, far exceeding the previous record for circulation half-life of 5.4 h. All these characteristics make these very powerful materials for in vivo applications, including drug delivery or imaging.

**Acknowledgements.**

**Supporting Information Available.** Materials and methods, as well as CMC's and dynamic light scattering spectra, AFM images and the PLE spectra of suspensions of SWNTs via (**2-3**), and the quenching of pyrene fluorescence on SWNTs and Gold NPs are available as supporting information. This material is available free of charge via the Internet at http://pubs.acs.org.

**Figure Captions**

**Figure 1.** a) PMHC$_{18}$-mPEG (**1**): poly(maleic anhydride-*alt*-1-octadecene)-poly(ethylene glycol) methyl ethers). b) γPGA-Py-mPEG (**2**): poly-(γ-glutamic acid)-Pyrine(30%)-poly(ethylene glycol) methyl ethers)(70%). c) γPGA-DSPE-mPEG (**3**): poly-(γ-glutamic acid)-phospholipid 1,2-distearoyl-*sn*-glycero-3-phosphoethanolamine (10%)-poly(ethylene glycol) methyl ethers)(60%).

**Figure 2.** a) Single-walled carbon nanotubes (SWNTs) coated with (**1**). b) AFM image of SWNTs with (**1**) coating. Non-uniform height along the nanotube is attributed to the presence of the polymer coating. c) UV-Vis-NIR absorption spectrum of SWNTs with (**1**) coating following excess polymer removal. Inset: Digital photograph of a SWNT suspension. d) Photoluminescence versus excitation (PLE) spectrum of SWNTs with (**1**) coating. Horizontal axis shows photoluminescence spectrum at different excitations along the vertical axis. Bright spots correspond to semiconducting SWNTs of different chiralities, demonstrating that the inherent NIR photoluminescent properties are retained with (**1**) coating.

**Figure 3.** a) UV-Vis absorption data of 20 nm gold nanoparticles with (**2**) (Inset: solution photo) or (**3**) coating in 50% fetal calf serum and as made (in water). Inset: TEM image of 20 nm Au NPs stabilized by (**2**). The UV-Vis spectra show the plasmon peak of gold nanoparticles at 530 nm in various coatings and media. b) 20 nm gold nanoparticles with to SH-mPEG (5kDa) coating with (left) and without (right) excess of SH-mPEG, showing aggregation following excess removal. c)

gold nanorods with (**2**) coating as made (in water, without excess of surfactant, solution photo) and in 50% fetal calf serum; gold nanorods with SH-mPEG (5kDa) coating (in water, without excess of surfactant). All Au NR suspensions were normalized in OD to the plasmon peak at 860 nm before removal of excess surfactant. Inset: TEM images of Au NRs stabilized by (**2**). The UV-Vis-NIR spectrum shows the transverse and longitudinal plasmon peaks of gold nanorods at 520 nm and 860 nm respectively.

**Figure 4.** Blood circulation data of SWNTs with different functionalizations in balb/c mice. (a) Blood circulation curves of (**1-3**) coated nanotubes compared with DSPE-branched-7kPEG coated nanotubes. The latter one was previously reported by our group[28]. Error bars were based on 3 mice per group at each time point (0.5h, 2h, 5h, 10h, 24h, 48h, 72h). (b) Blood circulation half-lives of different SWNTs obtained by first-order decay fitting of all data points. SWNTs with (**1**) and (**2**) coating exhibited drastically prolonged circulation half life compared with previous DSPE-PEG functionalized nanotubes.

**References**

1. Sinha, N.; Yeow, J. T. W., Carbon nanotubes for biomedical applications. *Ieee Transactions on Nanobioscience* **2005,** 4, (2), 180-195.
2. Han, G.; Ghosh, P.; Rotello, V. M., Functionalized gold nanoparticles for drug delivery. *Nanomedicine* **2007,** 2, (1), 113-123.
3. Kogan, M. J.; Olmedo, I.; Hosta, L.; Guerrero, A. R.; Cruz, L. J.; Albericio, F., Peptides and metallic nanoparticles for biomedical applications. *Nanomedicine* **2007,** 2, (3), 287-306.
4. Kam, N. W. S.; O'Connell, M.; Wisdom, J. A.; Dai, H. J., Carbon nanotubes as multifunctional biological transporters and near-infrared agents for selective cancer cell destruction. *Proceedings of the National Academy of Sciences of the United States of America* **2005,** 102, (33), 11600-11605.
5. Kam, N. W. S.; Liu, Z. A.; Dai, H. J., Carbon nanotubes as intracellular transporters for proteins and DNA: An investigation of the uptake mechanism and pathway. *Angewandte Chemie-International Edition* **2006,** 45, (4), 577-581.
6. Williams, K. A.; Veenhuizen, P. T. M.; de la Torre, B. G.; Eritja, R.; Dekker, C., Nanotechnology - Carbon nanotubes with DNA recognition. *Nature* **2002,** 420, (6917), 761-761.
7. Thaxton, C. S.; Georganopoulou, D. G.; Mirkin, C. A., Gold nanoparticle probes for the detection of nucleic acid targets. *Clinica Chimica Acta* **2006,** 363, (1-2), 120-126.

8. Maubach, G.; Csaki, A.; Born, D.; Fritzsche, W., Controlled positioning of a DNA molecule in an electrode setup based on self-assembly and microstructuring. *Nanotechnology* **2003,** 14, (5), 546-550.
9. Kryachko, E. S.; Remacle, F., Complexes of DNA bases and gold clusters Au-3 and Au-4 involving nonconventional N-H center dot center dot center dot Au hydrogen bonding. *Nano Letters* **2005,** 5, (4), 735-739.
10. Nath, N.; Chilkoti, A., Label free colorimetric biosensing using nanoparticles. *Journal of Fluorescence* **2004,** 14, (4), 377-389.
11. Nam, J. M.; Thaxton, C. S.; Mirkin, C. A., Nanoparticle-based bio-bar codes for the ultrasensitive detection of proteins. *Science* **2003,** 301, (5641), 1884-1886.
12. Chen, Z.; Tabakman, S. M.; Goodwin, A. P.; Kattah, M. G.; Daranciang, D.; Wang, X.; Zhang, G.; Li, X.; Liu, Z.; Utz, P. J.; Jiang, K.; Fan, S.; Dai, H., Protein microarrays with carbon nanotubes as multicolor Raman labels. *Nat Biotechnol* **2008,** 26, (11), 1285-92.
13. Liu, Z.; Cai, W. B.; He, L. N.; Nakayama, N.; Chen, K.; Sun, X. M.; Chen, X. Y.; Dai, H. J., In vivo biodistribution and highly efficient tumour targeting of carbon nanotubes in mice. *Nature Nanotechnology* **2007,** 2, (1), 47-52.
14. Adams, M. L.; Lavasanifar, A.; Kwon, G. S., Amphiphilic block copolymers for drug delivery. *Journal of Pharmaceutical Sciences* **2003,** 92, (7), 1343-1355.
15. Yang, S. T.; Fernando, K. A. S.; Liu, J. H.; Wang, J.; Sun, H. F.; Liu, Y. F.; Chen, M.; Huang, Y. P.; Wang, X.; Wang, H. F.; Sun, Y. P., Covalently PEGylated carbon nanotubes with stealth character in vivo. *Small* **2008,** 4, (7), 940-944.
16. Xiaoming Sun, Z. L., Kevin Welsher, Joshua Tucker Robinson, Andrew Goodwin, Sasa Zaric; Dai, a. H., Nano-Graphene Oxide for Cellular Imaging and Drug Delivery. *Nano Research* **2008,** 1, 203-212.
17. Lin Wang, W. Z., and Weihong Tan, Bioconjugated Silica Nanoparticles: Development and Applications. *Nano Research* **2008,** 1, 99-115.
18. Welsher, K.; Liu, Z.; Daranciang, D.; Dai, H., Selective probing and imaging of cells with single walled carbon nanotubes as near-infrared fluorescent molecules. *Nano Letters* **2008,** 8, (2), 586-590.
19. O'Connell, M. J.; Bachilo, S. M.; Huffman, C. B.; Moore, V. C.; Strano, M. S.; Haroz, E. H.; Rialon, K. L.; Boul, P. J.; Noon, W. H.; Kittrell, C.; Ma, J. P.; Hauge, R. H.; Weisman, R. B.; Smalley, R. E., Band gap fluorescence from individual single-walled carbon nanotubes. *Science* **2002,** 297, (5581), 593-596.
20. Richard, C.; Balavoine, F.; Schultz, P.; Ebbesen, T. W.; Mioskowski, C., Supramolecular self-assembly of lipid derivatives on carbon nanotubes. *Science* **2003,** 300, (5620), 775-778.
21. Numata, M.; Asai, M.; Kaneko, K.; Bae, A. H.; Hasegawa, T.; Sakurai, K.; Shinkai, S., Inclusion of cut and as-grown single-walled carbon nanotubes in the helical superstructure of schizophyllan and curdlan (ss-1,3-glucans). *Journal of the American Chemical Society* **2005,** 127, (16), 5875-5884.
22. Dieckmann, G. R.; Dalton, A. B.; Johnson, P. A.; Razal, J.; Chen, J.; Giordano, G. M.; Munoz, E.; Musselman, I. H.; Baughman, R. H.; Draper, R. K., Controlled assembly of carbon nanotubes by designed amphiphilic peptide helices. *Journal of the American Chemical Society* **2003,** 125, (7), 1770-1777.
23. Karajanagi, S. S.; Yang, H. C.; Asuri, P.; Sellitto, E.; Dordick, J. S.; Kane, R. S., Protein-assisted solubilization of single-walled carbon nanotubes. *Langmuir* **2006,** 22, (4), 1392-1395.
24. Zheng, M.; Jagota, A.; Semke, E. D.; Diner, B. A.; Mclean, R. S.; Lustig, S. R.; Richardson, R. E.; Tassi, N. G., DNA-assisted dispersion and separation of carbon nanotubes. *Nature Materials* **2003,** 2, (5), 338-342.

25. Zheng, M.; Jagota, A.; Strano, M. S.; Santos, A. P.; Barone, P.; Chou, S. G.; Diner, B. A.; Dresselhaus, M. S.; McLean, R. S.; Onoa, G. B.; Samsonidze, G. G.; Semke, E. D.; Usrey, M.; Walls, D. J., Structure-based carbon nanotube sorting by sequence-dependent DNA assembly. *Science* **2003,** 302, (5650), 1545-1548.
26. Sinani, V. A.; Gheith, M. K.; Yaroslavov, A. A.; Rakhnyanskaya, A. A.; Sun, K.; Mamedov, A. A.; Wicksted, J. P.; Kotov, N. A., Aqueous dispersions of single-wall and multiwall carbon nanotubes with designed amphiphilic polycations. *Journal of the American Chemical Society* **2005,** 127, (10), 3463-3472.
27. Chatterjee, T.; Yurekli, K.; Hadjiev, V. G.; Krishnamoorti, R., Single-walled carbon nanotube dispersions in poly(ethylene oxide). *Advanced Functional Materials* **2005,** 15, (11), 1832-1838.
28. Liu, Z.; Davis, C.; Cai, W. B.; He, L.; Chen, X. Y.; Dai, H. J., Circulation and long-term fate of functionalized, biocompatible single-walled carbon nanotubes in mice probed by Raman spectroscopy. *Proceedings of the National Academy of Sciences of the United States of America* **2008,** 105, (5), 1410-1415.
29. Cheng, Y.; Samia, A. C.; Meyers, J. D.; Panagopoulos, I.; Fei, B. W.; Burda, C., Highly efficient drug delivery with gold nanoparticle vectors for in vivo photodynamic therapy of cancer. *Journal of the American Chemical Society* **2008,** 130, (32), 10643-10647.
30. Huff, T. B.; Hansen, M. N.; Zhao, Y.; Cheng, J. X.; Wei, A., Controlling the cellular uptake of gold nanorods. *Langmuir* **2007,** 23, (4), 1596-1599.
31. Oyelere, A. K.; Chen, P. C.; Huang, X. H.; El-Sayed, I. H.; El-Sayed, M. A., Peptide-conjugated gold nanorods for nuclear targeting. *Bioconjugate Chemistry* **2007,** 18, (5), 1490-1497.
32. Sonnichsen, C.; Reinhard, B. M.; Liphardt, J.; Alivisatos, A. P., A molecular ruler based on plasmon coupling of single gold and silver nanoparticles. *Nature Biotechnology* **2005,** 23, (6), 741-745.
33. Cognet, L.; Tardin, C.; Boyer, D.; Choquet, D.; Tamarat, P.; Lounis, B., Single metallic nanoparticle imaging for protein detection in cells. *Proceedings of the National Academy of Sciences of the United States of America* **2003,** 100, (20), 11350-11355.
34. Schultz, S.; Smith, D. R.; Mock, J. J.; Schultz, D. A., Single-target molecule detection with nonbleaching multicolor optical immunolabels. *Proceedings of the National Academy of Sciences of the United States of America* **2000,** 97, (3), 996-1001.
35. Tong, L.; Zhao, Y.; Huff, T. B.; Hansen, M. N.; Wei, A.; Cheng, J. X., Gold nanorods mediate tumor cell death by compromising membrane integrity. *Advanced Materials* **2007,** 19, (20), 3136-+.
36. Paciotti, G. F.; Myer, L.; Weinreich, D.; Goia, D.; Pavel, N.; McLaughlin, R. E.; Tamarkin, L., Colloidal gold: A novel nanoparticle vector for tumor directed drug delivery. *Drug Delivery* **2004,** 11, (3), 169-183.
37. Gao, M. Y.; Kirstein, S.; Mohwald, H.; Rogach, A. L.; Kornowski, A.; Eychmuller, A.; Weller, H., Strongly photoluminescent CdTe nanocrystals by proper surface modification. *Journal of Physical Chemistry B* **1998,** 102, (43), 8360-8363.
38. Chan, W. C. W.; Nie, S. M., Quantum dot bioconjugates for ultrasensitive nonisotopic detection. *Science* **1998,** 281, (5385), 2016-2018.
39. Pinaud, F.; King, D.; Moore, H. P.; Weiss, S., Bioactivation and cell targeting of semiconductor CdSe/ZnS nanocrystals with phytochelatin-related peptides. *Journal of the American Chemical Society* **2004,** 126, (19), 6115-6123.
40. Kanaras, A. G.; Kamounah, F. S.; Schaumburg, K.; Kiely, C. J.; Brust, M., Thioalkylated tetraethylene glycol: a new ligand for water soluble monolayer protected gold clusters. *Chemical Communications* **2002**, (20), 2294-2295.

41.	Jiang, W.; Mardyani, S.; Fischer, H.; Chan, W. C. W., Design and characterization of lysine cross-linked mereapto-acid biocompatible quantum dots. *Chemistry of Materials* **2006,** 18, (4), 872-878.
42.	Dubertret, B.; Skourides, P.; Norris, D. J.; Noireaux, V.; Brivanlou, A. H.; Libchaber, A., In vivo imaging of quantum dots encapsulated in phospholipid micelles. *Science* **2002,** 298, (5599), 1759-1762.
43.	Fan, H. Y.; Leve, E. W.; Scullin, C.; Gabaldon, J.; Tallant, D.; Bunge, S.; Boyle, T.; Wilson, M. C.; Brinker, C. J., Surfactant-assisted synthesis of water-soluble and biocompatible semiconductor quantum dot micelles. *Nano Letters* **2005,** 5, (4), 645-648.
44.	Mulder, W. J. M.; Koole, R.; Brandwijk, R. J.; Storm, G.; Chin, P. T. K.; Strijkers, G. J.; Donega, C. D.; Nicolay, K.; Griffioen, A. W., Quantum dots with a paramagnetic coating as a bimodal molecular imaging probe. *Nano Letters* **2006,** 6, (1), 1-6.
45.	LizMarzan, L. M.; Giersig, M.; Mulvaney, P., Homogeneous silica coating of vitreophobic colloids. *Chemical Communications* **1996**, (6), 731-732.
46.	Bruchez, M.; Moronne, M.; Gin, P.; Weiss, S.; Alivisatos, A. P., Semiconductor nanocrystals as fluorescent biological labels. *Science* **1998,** 281, (5385), 2013-2016.
47.	Gerion, D.; Pinaud, F.; Williams, S. C.; Parak, W. J.; Zanchet, D.; Weiss, S.; Alivisatos, A. P., Synthesis and properties of biocompatible water-soluble silica-coated CdSe/ZnS semiconductor quantum dots. *Journal of Physical Chemistry B* **2001,** 105, (37), 8861-8871.
48.	Selvan, S. T.; Tan, T. T.; Ying, J. Y., Robust, non-cytotoxic, silica-coated CdSe quantum dots with efficient photoluminescence. *Advanced Materials* **2005,** 17, (13), 1620-+.
49.	Zhelev, Z.; Ohba, H.; Bakalova, R., Single quantum dot-micelles coated with silica shell as potentially non-cytotoxic fluorescent cell tracers. *Journal of the American Chemical Society* **2006,** 128, (19), 6324-6325.
50.	Sohn, B. H.; Choi, J. M.; Yoo, S. I.; Yun, S. H.; Zin, W. C.; Jung, J. C.; Kanehara, M.; Hirata, T.; Teranishi, T., Directed self-assembly of two kinds of nanoparticles utilizing monolayer films of diblock copolymer micelles. *Journal of the American Chemical Society* **2003,** 125, (21), 6368-6369.
51.	Porta, F.; Prati, L.; Rossi, M.; Scari, G., Synthesis of Au(0) nanoparticles from W/O microemulsions. *Colloids and Surfaces a-Physicochemical and Engineering Aspects* **2002,** 211, (1), 43-48.
52.	Markowitz, M. A.; Dunn, D. N.; Chow, G. M.; Zhang, J., The effect of membrane charge on gold nanoparticle synthesis via surfactant membranes. *Journal of Colloid and Interface Science* **1999,** 210, (1), 73-85.
53.	Chen, R. J.; Zhang, Y. G.; Wang, D. W.; Dai, H. J., Noncovalent sidewall functionalization of single-walled carbon nanotubes for protein immobilization. *Journal of the American Chemical Society* **2001,** 123, (16), 3838-3839.
54.	Liu, W. H.; Choi, H. S.; Zimmer, J. P.; Tanaka, E.; Frangioni, J. V.; Bawendi, M., Compact cysteine-coated CdSe(ZnCdS) quantum dots for in vivo applications. *Journal of the American Chemical Society* **2007,** 129, (47), 14530-+.
55.	Ipe, B. I.; Thomas, K. G., Investigations on nanoparticle-chromophore and interchromophore interactions in pyrene-capped gold nanoparticles. *Journal of Physical Chemistry B* **2004,** 108, (35), 13265-13272.
56.	Wang, D.; Ji, W. X.; Li, Z. C.; Chen, L. W., A biomimetic "polysoap" for single-walled carbon nanotube dispersion. *Journal of the American Chemical Society* **2006,** 128, (20), 6556-6557.
57.	Nakayama-Ratchford, N.; Bangsaruntip, S.; Sun, X. M.; Welsher, K.; Dai, H. J., Noncovalent functionalization of carbon nanotubes by fluorescein-polyethylene glycol: Supramolecular conjugates with pH-dependent absorbance and fluorescence. *Journal of the American Chemical Society* **2007,** 129, (9), 2448-+.

58. Ashiuchi, M.; Nawa, C.; Kamei, T.; Song, J. S.; Hong, S. P.; Sung, M. H.; Soda, K.; Yagi, T.; Misono, H., Physiological and biochemical characteristics of poly-g-glutamate synthetase complex of Bacillus subtilis (vol 268, pg 5321, 2000). *European Journal of Biochemistry* **2001,** 268, (22), 6003-6003.
59. Parveen, S.; Sahoo, S. K., Nanomedicine - Clinical applications of polyethylene glycol conjugated proteins and drugs. *Clinical Pharmacokinetics* **2006,** 45, (10), 965-988.

Figure 1

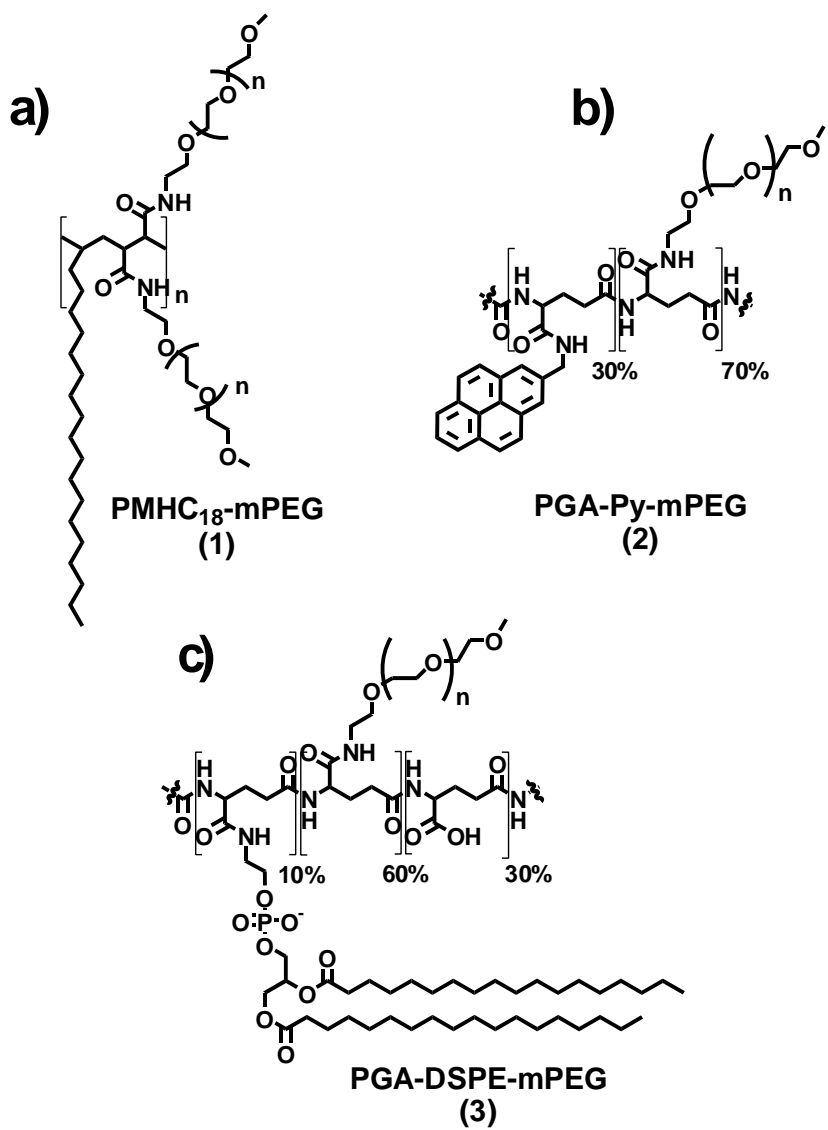

Figure 2

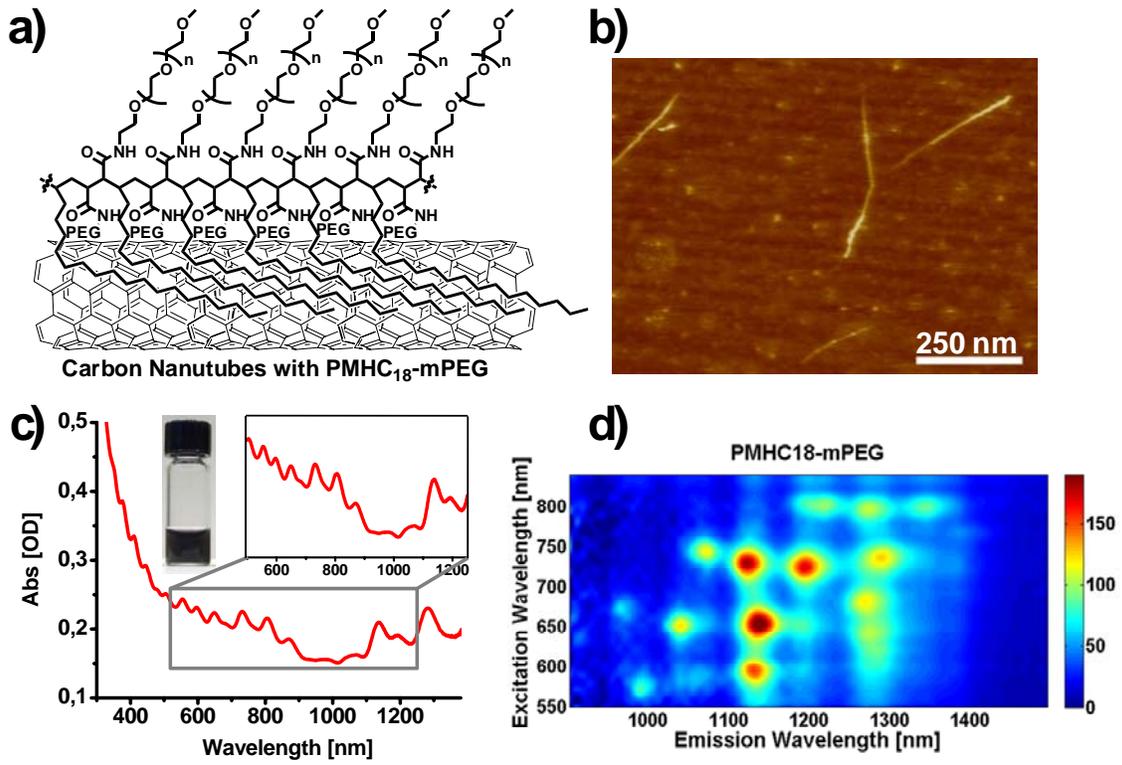

Figure 3

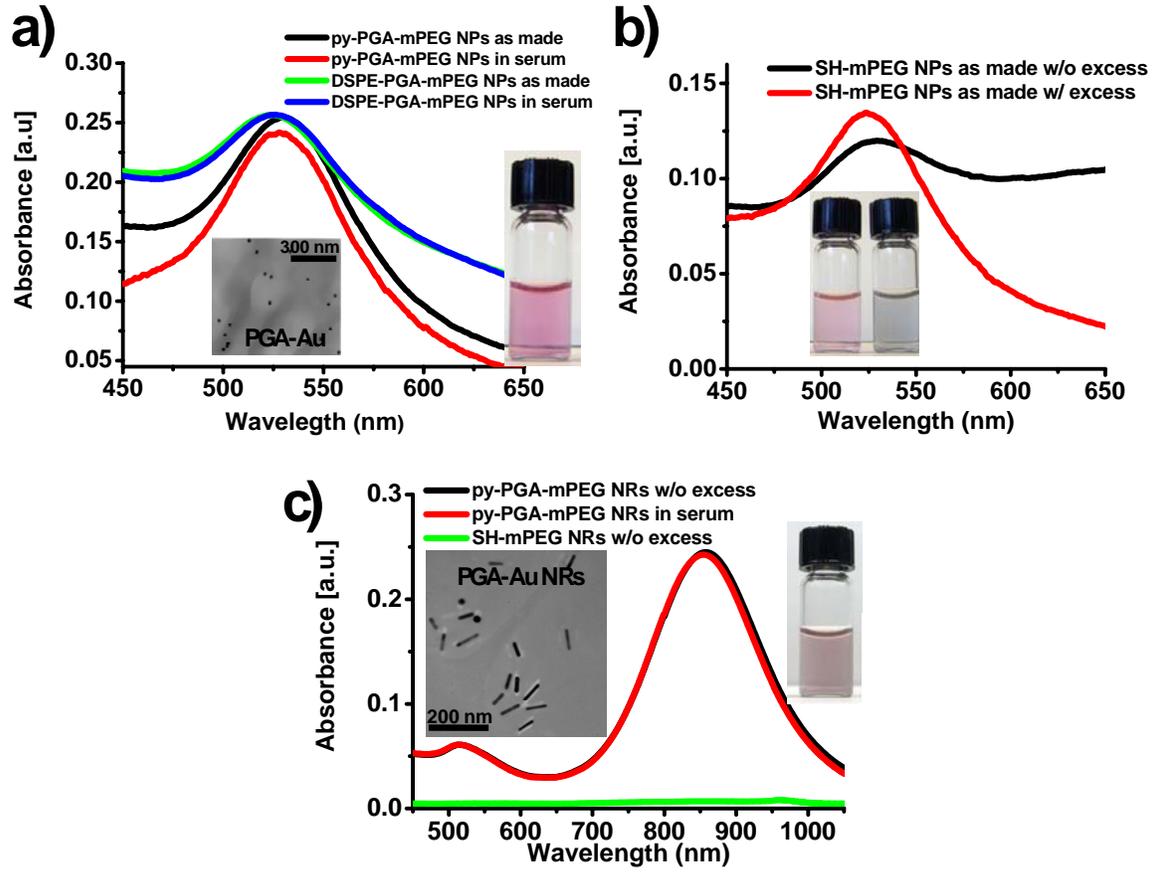

Figure 4

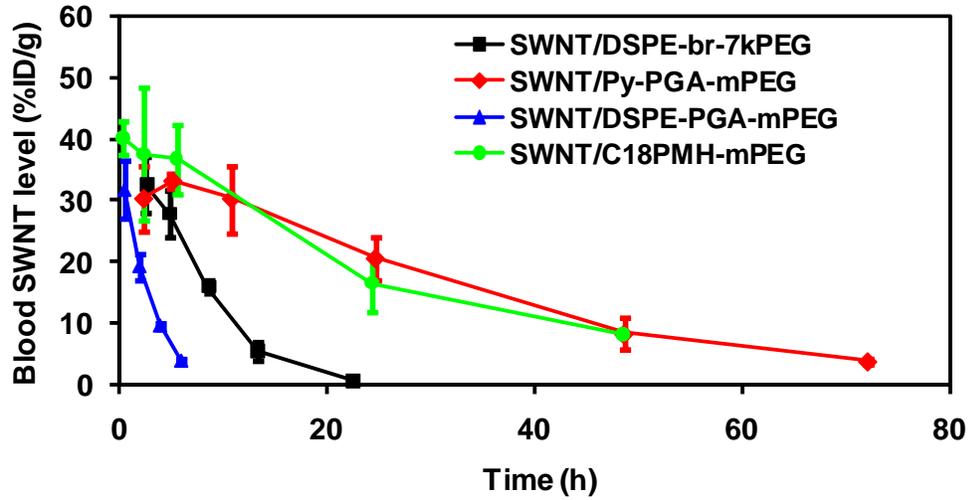

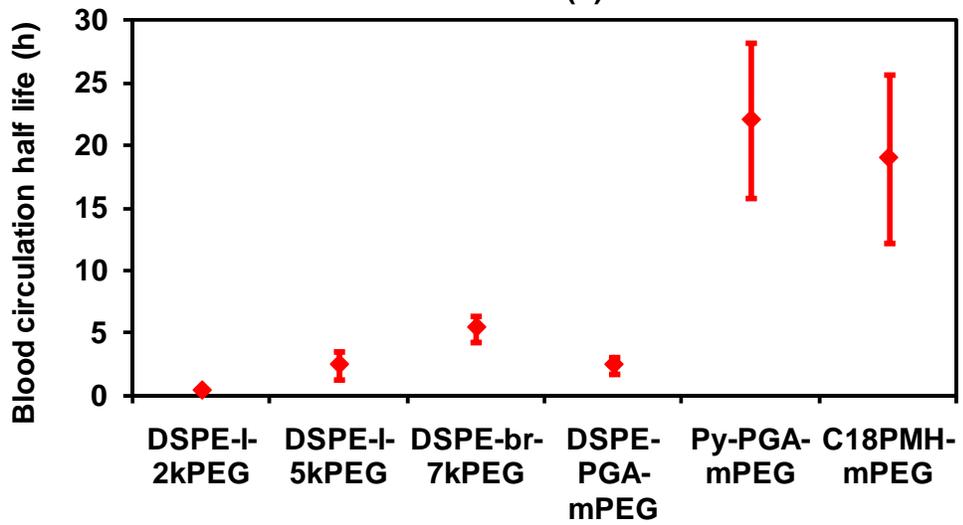

**Table of Contents**

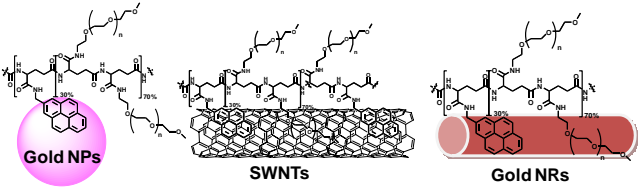